\documentclass[aps,prl,twocolumn,showpacs]{revtex4-1}

\usepackage{amsmath}
\usepackage{amssymb}
\usepackage{graphicx}
\usepackage{epstopdf}

\begin{document}
\title{Topological Effects on Quantum Phase Slips in Superfluid Spin Transport}

\author{Se Kwon Kim}
\affiliation{
	Department of Physics and Astronomy,
	University of California,
	Los Angeles, California 90095, USA
}

\author{Yaroslav Tserkovnyak}
\affiliation{
	Department of Physics and Astronomy,
	University of California,
	Los Angeles, California 90095, USA
}

\date{\today}

\begin{abstract}
We theoretically investigate effects of quantum fluctuations on superfluid spin transport through easy-plane quantum antiferromagnetic spin chains in the large-spin limit. Quantum fluctuations give rise to decaying of spin supercurrent by unwinding the magnetic order parameter within the easy plane, which is referred to as phase slips. We show that the topological term in the nonlinear sigma model for the spin chains qualitatively differentiates decaying rate of the spin supercurrent between integer spin and half-odd-integer spin chains. An experimental setup for a magnetoelectric circuit is proposed, in which the dependence of the decaying rate on constituent spins can be verified by measuring nonlocal magnetoresistance.
\end{abstract}


\maketitle

\emph{Introduction.}|Quantum fluctuations have important effects on physical properties of low dimensional systems as exemplified by the Mermin-Wagner theorem that excludes continuous-symmetry-breaking order in one-dimensional systems at all temperatures \cite{MerminPRL1966}. One-dimensional quantum magnetism has thus been a natural playground to seek and study exotic states that deny classical descriptions \cite{*[][{, and references therein.}] AffleckJPCM1989, *[][{, and references therein.}] Mikeska2004}. A prototypical example showing importance of quantum effects is provided by Heisenberg antiferromagnetic spin chains. For isotropic spin-$s$ chains, Haldane suggested in 1983 \cite{HaldanePL1983, *HaldanePRL1983} that integer-$s$ chains have disordered ground states with gapped excitations unlike half-odd-integer-$s$ chains having gapless excitations \cite{LiebAP1961, *AffleckLMP1986, *ShankarNPB1990}. The existence of the gap has been experimentally confirmed for $s = 1$ \cite{BuyersPRL1986, *RenardEPL1987}.

\begin{figure}
\includegraphics[width=\columnwidth]{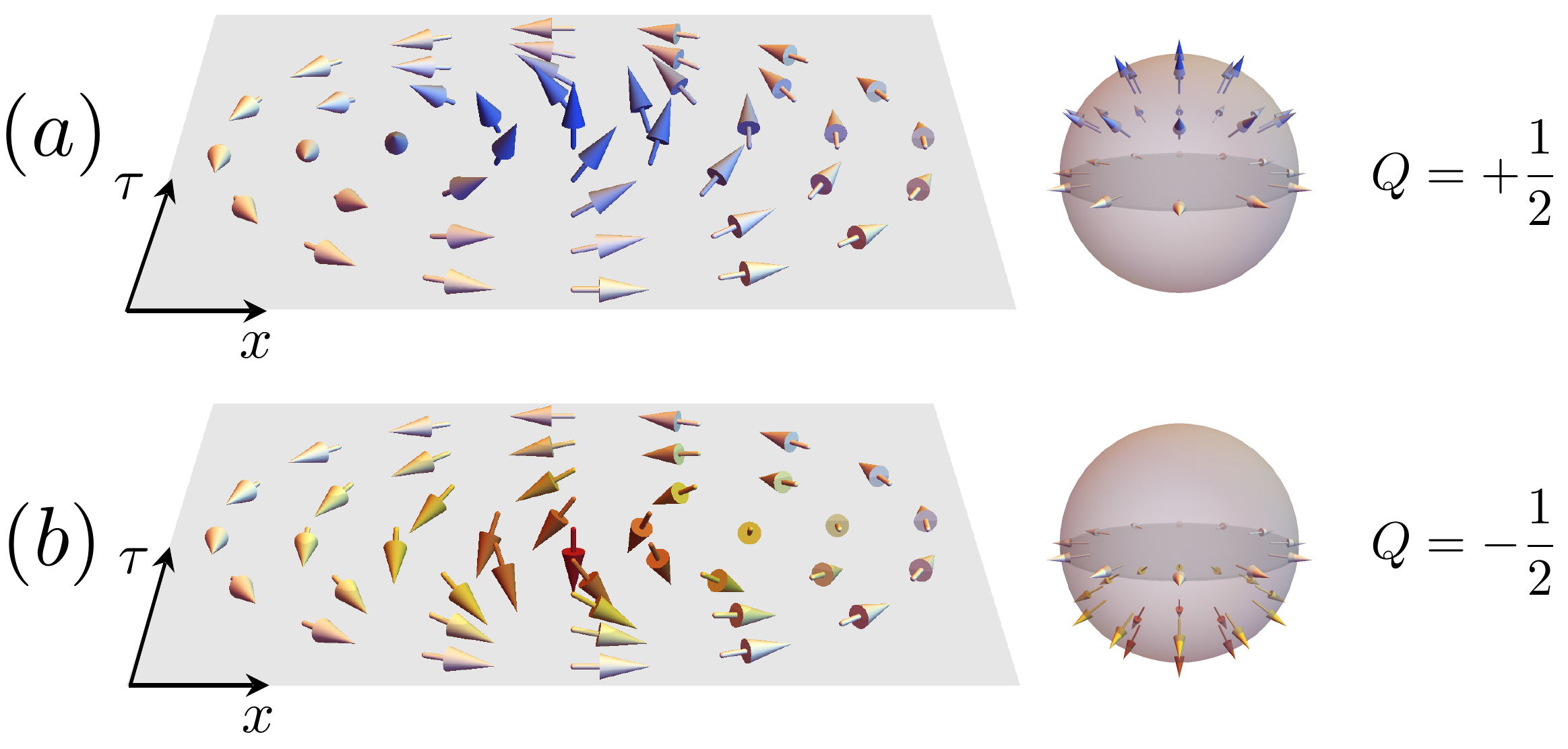}
\caption{(color online) Vortex configurations of the local N\'eel order parameter in the Euclidean spacetime $(x, \tau)$ with skyrmion charges (a) $Q = 1/2$ and (b) $Q = -1/2$.}
\label{fig:fig1}
\end{figure}

By considering anisotropic antiferromagnetic spin chains in the large-$s$ limit, \textcite{AffleckPRL1986-1} was able to attribute this distinction between integer and half-odd-integer spin chains to the topological term in the O(3) nonlinear sigma model that describes the dynamics of the local N\'eel order parameter \cite{MikeskaJPC1980, HaldanePL1983, FradkinPRB1988}. For sufficiently large $s$, easy-plane spin-$s$ chains are in the gapless XY phase, where order-destroying excitations are vortices of the order parameter in the two-dimensional Euclidean spacetime. It is the skyrmion charge $Q$ of a vortex, quantifying how many times the order parameter wraps the unit sphere, that serves as the topological charge in the nonlinear sigma model. Figure~\ref{fig:fig1} illustrates vortices with minimum nonzero skyrmion charges $Q = \pm 1/2$, which are often referred to as merons \cite{GrossNPB1978}. Only for half-odd-integer spin chains, the topological term creates destructive interference between vortices and, thereby, suppress effects of quantum fluctuations \cite{AffleckJPCM1989, IvanovPRB1998}.

Superfluid spin transport, a spin analog of an electrical supercurrent, has been proposed in magnets with easy-plane anisotropy, where the direction of the local magnetic order within the easy plane plays a role of the phase of superfluid order parameter \cite{SoninJETP1978, *SoninAP2010, KonigPRL2002, *ChenPRB2014, ChenPRB2014-2, *ChenPRL2015, TakeiPRL2014, *TakeiPRB2014}. Spin supercurrent therein is sustained by spiraling texture of the magnetic order, being proportional the gradient of in-plane components of the order parameter. Under the guideline of established theories for resistance in superconducting wires \cite{*[][{, and references therein.}] HalperinIJMPB2010}, we have recently investigated intrinsic thermal resistance in one-dimensional superfluid spin transport, which arises via thermally-activated phase slips \cite{LittlePR1967, *LangerPR1967, *McCumberPRB1970} that unwind the phase by lifting the magnetic order off the easy plane \cite{KimarXiv2015}. At sufficiently low temperatures, however, resistance is mainly induced by quantum fluctuations via quantum phase slips (QPS) \cite{*[][{, and references therein.}] GiordanoP1994, ZaikinPRL1997}. QPS in superconducting wires correspond to vortices of the phase of the order parameter in the Euclidean spacetime. Likewise, QPS in one-dimensional spin superfluidity correspond to vortices of the magnetic order parameter. Then, there arises a natural question how the topological term distinguishes integer-$s$ chains and half-odd-integer-$s$ chains in QPS-induced resistance of superfluid spin transport.

\begin{figure}
\includegraphics[width=\columnwidth]{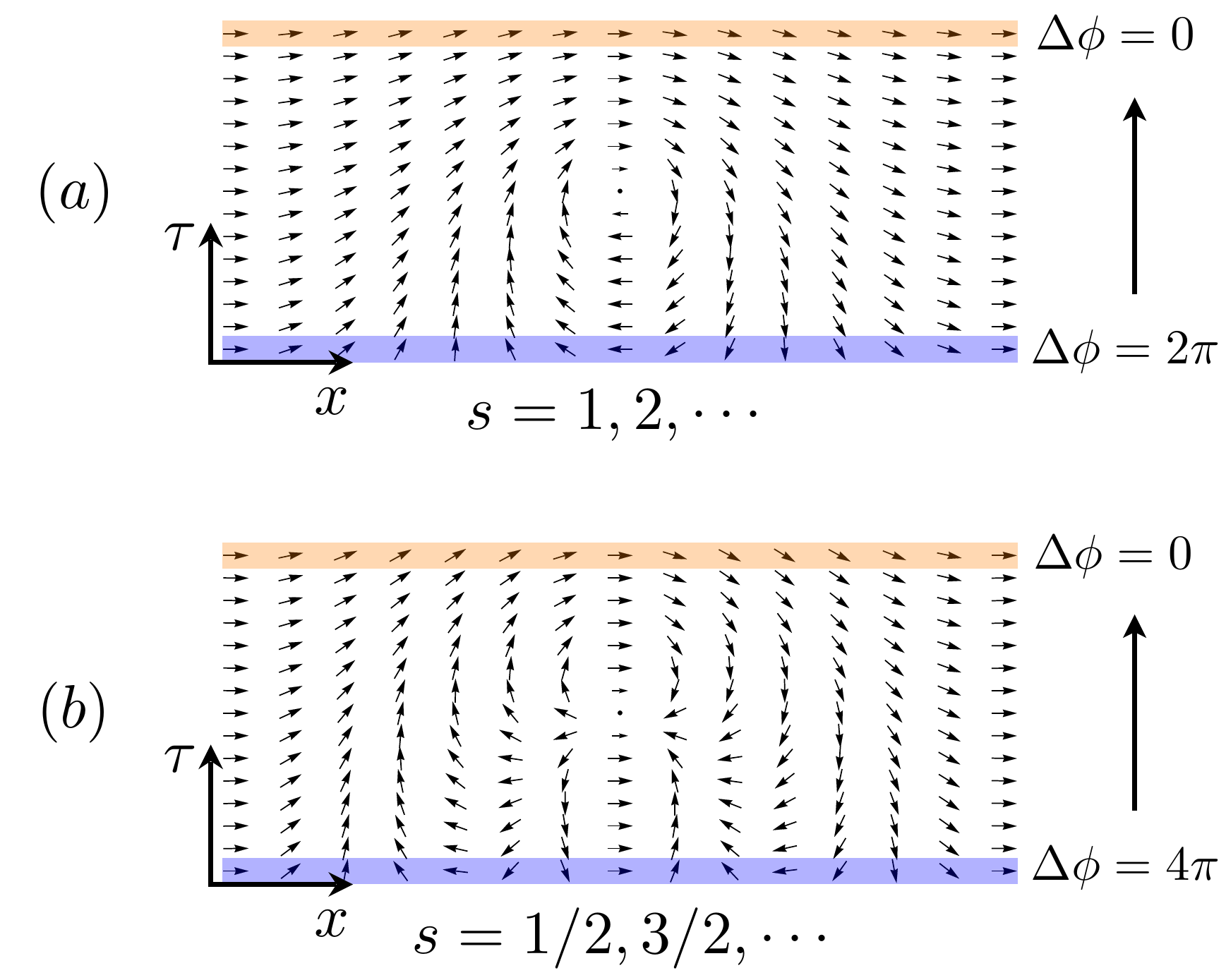}
\caption{(color online) Decaying of the spin current, which is proportional to the winding number $\Delta \phi$, via QPS with skyrmion charges (a) $Q = 1/2$ and (b) $Q = 1$. For half-odd-integer spin chains, $2 \pi$ phase slips are prohibited by destructive interference between QPS with skyrmion charges $Q = \pm 1/2$. See the main text for detailed discussions.}
\label{fig:fig2}
\end{figure}

To answer the question, we theoretically study QPS in superfluid spin transport through easy-plane quantum antiferromagnetic spin chains in this Letter. For integer $s$, the topological term is inactive, and resistance arises due to QPS of skyrmion charges $Q = \pm 1/2$ that change winding number by $2 \pi$. For half-odd-integer $s$, those QPS are completely suppressed due to destructive interferences. Instead, QPS of twice-larger skyrmion charges $Q = \pm 1$, give rise to resistance by unwinding the phase by $4 \pi$. See Fig.~\ref{fig:fig2} for illustrations of QPS. Resistance in superfluid spin transport can be characterized by decaying rate of the spin supercurrent, $\kappa(I, T)$, which is a function of the spin supercurrent $I$ and the ambient temperature $T$. Our main finding is qualitative difference of the decaying rate between integer-$s$ and half-odd-integer-$s$ spin chains, which can be summarized as
\begin{equation}
\kappa(I, T) \propto
\begin{cases}
	T^{2 \mu -3} \quad & \text{for } I \ll T \\
	I^{2 \mu -3} \quad & \text{for } T \ll I
\end{cases} 
\, ,
\end{equation}
where $\mu$ is given by
\begin{equation}
\mu = 
\begin{cases}
	\pi s / 2 \quad 	&\text{ for integer } s \\
	2 \pi s \quad	&\text{ for half-odd-integer } s
\end{cases} \, .
\label{eq:mu}
\end{equation}
The exponent $\mu$ parametrizes the strength of interaction between QPS, which is proportional to the square of their skyrmion charges; $\mu$ is thus four times larger for half-odd-integer $s$ than for integer $s$. This contrast between integer-$s$ and half-odd-integer-$s$ spin chains can be demonstrated by measuring voltage- or temperature-dependence of electrical resistance of the magnetoelectric circuit in Ref.~\cite{TakeiPRL2015}, which has been proposed for probing superfluid spin transport, using a quasi-one-dimensional antiferromagnet, e.g., (CH$_3$)$_4$NMnCl$_3$ ($s = 5/2$) \cite{HutchingsPRB1972, *BoucherJMMM1979, *FluggenSSC1983} as a spin transport channel. 

\emph{Model.}|We consider an anisotropic Heisenberg antiferromagnetic spin-$s$ chain that can be described by the Hamiltonian 
\begin{equation}
H = J \sum_n \left[ \mathbf{S}_n \cdot \mathbf{S}_{n+1} - a S_n^z S_{n+1}^z + b (S_n^z)^2 \right]
\label{eq:H}
\end{equation}
with $\mathbf{S}_n^2 = s (s+1)$, where small positive constants $a \ll 1$ and $b \ll 1$ parametrize the anisotropy. In the large-$s$ limit, neighboring spins are mostly antiparallel, $\mathbf{S}_n \approx - \mathbf{S}_{n+1}$ in low-energy states, and long wavelength dynamics of the chain can be understood in terms of the slowly varying unit vector $\mathbf{n} = (\mathbf{S}_{2 n} - \mathbf{S}_{2 n + 1})/s$ in the direction of the local N\'eel order parameter. The dynamics of the field $\mathbf{n}$ follows the nonlinear sigma model \cite{MikeskaJPC1980, HaldanePL1983, AffleckPRL1986-1, FradkinPRB1988} with Euclidean action $S = i \theta Q + S_0$ (in units of $\hbar$), where $\theta \equiv 2 \pi s$ is referred to as the topological angle. Here,
\begin{equation}
Q \equiv \frac{1}{4 \pi} \int dx \int_0^{\hbar \beta} d\tau \, \mathbf{n} \cdot (\partial_{x} \mathbf{n} \times \partial_{\tau} \mathbf{n})
\end{equation}
is the skyrmion charge of $\mathbf{n}$ that measures how many times $\mathbf{n}(x, \tau)$ wraps the unit sphere as the space and imaginary-time coordinates, $x$ and $\tau$, vary, and, is thus topological. A nontopological part of the action is given by
\begin{equation}
S_0 	= \frac{1}{2 g} \int dx \int_0^{\hbar \beta c} d (c \tau) \left[ \frac{(\partial_\tau \mathbf{n})^2}{c^2} + (\partial_x \mathbf{n})^2 + \frac{n_z^2}{\lambda^2} \right] \, ,
\label{eq:S0}
\end{equation}
where $c \equiv 2 J s d / \hbar$ serves as a speed of ``light" for the theory, $d$ is the lattice constant, and $\lambda \equiv d / \sqrt{2 (a + b)}$ is a characteristic length scale formed by the anisotropy. Ultraviolet and infrared cutoffs are provided by the length scales, $d$ and $\lambda$, respectively. Here, $g \equiv 2/s$ is the dimensionless coupling constant, which sets the quantum ``temperature" governing magnitude of quantum fluctuations \cite{HaldanePRL1983}.

The corresponding partition function is given by
\begin{equation}
\mathcal{Z} = \int \mathcal{D} \mathbf{n}(x, \tau) \delta(\mathbf{n}^2 - 1) \exp(- i \theta Q - S_0) \, .
\end{equation}
We consider the field $\mathbf{n}$ that are periodic in the imaginary time $\tau$, $\mathbf{n} (x, \tau) = \mathbf{n} (x, \tau + \hbar \beta)$. The partition function $\mathcal{Z}$ is then a periodic function of the topological angle $\theta$. For integer and half-odd-integer $s$, therefore, we can effectively set $\theta = 0$ and $\theta = \pi$, respectively \cite{AffleckJPCM1989}.

\emph{Spin superfluidity.}|To discuss spin superfluidity associated with the invariance of the action under spin rotations about the $z$ axis, it is convenient to parametrize $\mathbf{n}$ in spherical coordinates, $\theta$ and $\phi$, defined by $\mathbf{n} = (\sin \theta \cos \phi, \sin \theta \sin \phi, \cos \theta)$. Steady states carrying uniform spin current are given by time-independent solutions to the Euler-Lagrange equation of the action:
\begin{equation}
\theta(x) = \pi / 2 \, , \quad \phi(x) = \phi_0 + k x \quad \left( |k| < \lambda^{-1} \right) \, ,
\label{eq:ss}
\end{equation}
with $\phi_0$ an arbitrary reference angle \cite{SoninJETP1978}. The infrared cutoff $\lambda^{-1}$ sets a critical current for stable superfluid spin transport. In these steady states, the uniform spin supercurrent, $I = - J s^2 k d$, is sustained by spiraling texture of $\mathbf{n}$ within the easy plane, which is analogous to the electrical supercurrent maintained by gradient of the phase of the superconducting order parameter. When the chain is long enough, $L \gg \lambda$, which we assume henceforth, actual boundary conditions at the ends of the chain are not important. Imposing periodic boundary conditions on the order parameter, $\mathbf{n} (x = 0, \tau) = \mathbf{n} (x = L, \tau)$, quantizes allowed spin supercurrent, $k_\nu = 2 \pi \nu / L$, where $\nu = \Delta \phi / 2 \pi$ is the winding number of $\mathbf{n}$ in the easy plane. 

\emph{QPS in spin superfluidity.}|The spin supercurrent in a closed chain can be indefinitely maintained if there are no fluctuations. Finite dissipation, however, arises due to thermal and quantum fluctuations, which provide transition channels between steady states with different winding numbers $\nu \neq \nu'$ \cite{KimarXiv2015}. Such events changing winding numbers are referred to as phase slips. In this Letter, we are interested in QPS, which dominate thermally-activated phase slips at sufficiently low temperatures. We, heretofore, consider situations where a temperature is much smaller than the characteristic energy scale of the spin chain, $T \ll \hbar c / \lambda$.

QPS are vortex configurations of $\mathbf{n}$ in the two-dimensional Euclidean spacetime \cite{HalperinIJMPB2010}. For a single vortex centered at the origin, which is a saddle point of the action $S_0$, the azimuthal angle is given by
\begin{equation}
\phi_q (x, \tau) = \phi_0 + q \arctan (c \tau / x) \, ,
\end{equation}
where nonzero integer $q$ is a vorticity of the vortex. The polar angle is given by a function $\theta (r)$ of the radial distance $r \equiv \sqrt{x^2 + c^2 \tau^2}$, which solves a differential equation, $d^2 \theta / dr^2 + (1/r) d\theta / dr = \sin \theta \cos \theta (1 / \lambda^2 - q^2 / r^2)$ with boundary conditions, $\theta(0) = (1 - p) \pi / 2$ and $\theta(r \to \infty) = \pi/2$ \cite{IvanovPRL1994}. The order parameter $\mathbf{n}$ is substantially off the easy plane only within the disk $r \lesssim \lambda$, defining a core of the vortex. At the center of the vortex, the order parameter points either the north pole $p = +1$ or the south pole $p = -1$, which is referred to as a polarity of the vortex. Vortex solutions are characterized by vorticity $q$ and polarity $p$, which is related to the skyrmion charge as $Q = p q / 2$ \cite{TretiakovPRB2007}. See Fig.~\ref{fig:fig1} for illustrations of vortices with $Q = \pm 1/2$.

Let us now consider a dilute gas of $n$ QPS in the background of small spin current $k \ll \lambda^{-1}$. The gas of QPS must be vorticity-neutral $\sum_{i} q_i = 0$ to meet the periodic boundary conditions $\mathbf{n} (x, \tau) = \mathbf{n} (x, \tau + \hbar \beta) = \mathbf{n} (x + L, \tau)$. Substituting the saddle point solution, $\phi = k x + \sum_i \phi_{q_i} (x - x_i, \tau - \tau_i)$ and the corresponding $\theta(x, \tau; \{ p_i \})$, into the action, we find
\begin{eqnarray}
S	&=& i \theta \sum_i {p_i q_i} / 2 + S_0 \, , \label{eq:S} \\
S_0	&=& \sum_i S_\text{core} (q_i) - (2 \pi / g) \sum_{i < j} q_i q_j \ln ( d_{ij} / \lambda ) \, \nonumber \\
	&& + (2 \pi / g) c k \sum_i q_i \tau_i \, ,
\end{eqnarray}
where $d_{ij} = \sqrt{(x_i - x_j)^2 + c^2 (\tau_i - \tau_j)^2} \gg \lambda$ is the distance between QPS \footnote{With the background spin current $k$, $S_\text{core}$ acquires a new contribution that is order of $k^2$. This change of $S_\text{core}$ does not affect qualitative behavior of QPS, and is thus ignored.}. The nontopological part of the action $S_0$ consists of three terms. The first term is the contribution from vortex cores to the action, which can be estimated as $S_\text{core} \approx \pi / g$. The second term is logarithmic interaction between QPS. The third term is coupling of QPS to the spin current $k$.

The topological term $i \theta \sum_i p_i q_i / 2$ depends on polarities $\{ p_i \}$ of QPS, whereas the nontoplogical term $S_0$ does not. For fixed vorticity configuration $\{ q_i \}$, the partition function is summed over two possible polarities for each QPS, $p_i = \pm 1$, which results in
\begin{equation}
\mathcal{Z} \propto \left[ \prod_{i} \cos \left( \frac{\theta q_i}{2} \right) \right] e^{- S_0( \{ q_i \} )} \, . 
\end{equation}
As pointed out by \textcite{AffleckPRL1986-1} to explain the gapped Haldane phase of integer-$s$ spin chains \cite{HaldanePL1983, AffleckPRL1986-1}, the prefactor of the partition function distinguishes integer and half-odd-integer $s$. For integer $s$, the topological angle is zero $\theta = 0$, and thus the prefactor is $1$. Half-odd-integer $s$, however, yields $\theta = \pi$, and the prefactor vanishes when any of vorticities $\{ q_i \}$ is odd. This destructive interference between QPS with odd vorticities can be effectively captured by setting an elementary vorticity of QPS to $2$. Let us use the symbol $q_0$ to denote an elementary vorticity; $q_0 = 1$ and $q_0 = 2$ for integer and half-odd-integer $s$, respectively. Low-energy dynamics of the order parameter will be dominantly affected by QPS with the elementary vorticity. We therefore focus on a gas of such QPS henceforth, which is described by the effective action:
\begin{equation}
S_\text{eff} = n S_\text{core} - 2 \mu \sum_{i < j} \tilde{q}_i \tilde{q}_j \ln( d_{ij} / \lambda ) + \sigma \sum_i \tilde{q}_i \tau_i \, ,
\label{eq:S-eff}
\end{equation}
where $\mu \equiv \pi q_0^2 / g$ sets the interaction strength between QPS, $\sigma \equiv 2 \pi q_0 c k / g$ is rescaled spin current, and $\tilde{q}_i \equiv q_i / q_0 = \pm 1$ is a rescaled vorticity. The effective action $S_\text{eff}$ without the last term has been discussed to study phase diagram of spin chains, e.g, in Ref.~\cite{AffleckJPCM1989}.

\emph{Tapping into superconducting wires.}|Owing to the formal equivalence of the action $S_\text{eff}$ to the action for a gas of QPS in a superconducting wire, specifically Eq.~(4) in Ref.~\cite{ZaikinPRL1997}, we can adopt the results for superconductivity to our case of spin superfluidity. First of all, there is a superfluid-to-insulator phase transition at the critical interaction strength $\mu^*$ in the absence of the spin current, $\sigma = 0$. For $\mu > \mu^*$, QPS attract strongly and form bound pairs, keeping spin superfluidity intact. As $\mu$ decreases below $\mu^*$, QPS condense and destroy spin superfluidity, driving the system to the insulating phase. These insulating and superfluid phases are, respectively, the gapped Haldane and the gapless XY phases of anisotropic spin chains \cite{AffleckJPCM1989}. The condition for being in the superfluid phase is $\mu > \mu^* \approx 2$ \cite{KosterlitzJPC1973, *KosterlitzJPC1974, ZaikinPRL1997}, which corresponds to $s\ge 2$ and $s \ge 1/2$ for integer and half-odd-integer $s$, respectively \footnote{To keep the discussion on the phase transition simple, we used the bare value of the critical strength, $\mu^* \approx 2$, which does not take its renormalization into account.}.

Secondly, QPS rates have been derived for a superconducting wire in Ref.~\cite{ZaikinPRL1997} by following the Langer's theory for the decay of metastable states \cite{LangerAP1967, *WeissPRB1987}. By adopting the result into spin superfluidity, we can find the average decaying rate $\kappa(I, T)$ of the winding number, $\dot{\nu} = - \kappa \nu$, as a function of the spin current $I$ and the ambient temperature $T$:
\begin{equation}
\begin{split}
\kappa(I, T) 	& = z^2 \omega_0 (T / \hbar \omega_0)^{2 \mu - 2} \mathcal{F} (I/T) \\
\mathcal{F} (\xi)	& \equiv C \sinh(\xi / 2) \left| \Gamma ( \mu - 1/2 + i \xi / 2 \pi ) \right|^2
\end{split} \, , 
\label{eq:kappa1}
\end{equation}
where $z \equiv \exp(- S_\text{core})$ is the fugacity of QPS, $\omega_0 \equiv c / \lambda$ is the characteristic frequency of the spin chain, and $C \equiv 8 \pi^{3/2} (2 \pi)^{2 \mu -2} \Gamma(\mu - 1/2) / \Gamma(\mu) \Gamma(2 \mu - 1)$ is a numerical constant \footnote{Equation~(\ref{eq:kappa1}) for $\kappa$ is obtained without taking into account quantum fluctuations about vortex solutions, which may add an additional dimensionless factor that depends neither on $I$ nor $T$ \cite{ZaikinPRL1997}.}. The expression for $\kappa(I, T)$ is simplified when one parameter dominates the other \footnote{There are a few distinct ways to obtain the resistivity of one-dimensional superfluid (see e.g., Refs.~\cite{GiamarchiPRB1992, DanshitaPRA2012}), which yield different expressions for the decaying rate $\kappa(I, T)$. In two limits, $I \ll T$ and $T \ll I$, however, the dependence of $\kappa(I, T)$ on the spin current $I$ and the temperature $T$ is independent of employed methods.};
\begin{equation}
\kappa(I, T) \propto
\begin{cases}
	z^2 \omega_0 (T / \hbar \omega_0)^{2 \mu -3} \quad & \text{for } I \ll T \\
	z^2 \omega_0 (I / \hbar \omega_0)^{2 \mu -3} \quad & \text{for } T \ll I
\end{cases} 
\, .
\label{eq:kappa2}
\end{equation}
The interaction strength $\mu$ is given in Eq.~(\ref{eq:mu}). This is our main result: the topological term in the nonlinear sigma model for the Heisenberg antiferromagnetic spin chains distinguishes between integer and half-odd-integer spins in the decaying rate of the spin supercurrent. Figure~\ref{fig:fig2} illustrates decaying of the spin current via QPS.

To see such quantum effects, we should work at sufficiently low temperatures, where quantum fluctuations dominate thermal ones. The crossover temperature $T^*$ can be estimated by matching the classical phase-slip energy barrier (divided by $T$) \cite{KimarXiv2015} to the action of two noninteracting QPS \cite{ZaikinPRL1997}, $2 \hbar c / \lambda T^* \approx 2 S_\text{core}$. Using $S_\text{core} \approx \pi / g$ yields $T^* \approx 2 \hbar c / \pi s \lambda$.

\begin{figure}
\includegraphics[width=\columnwidth]{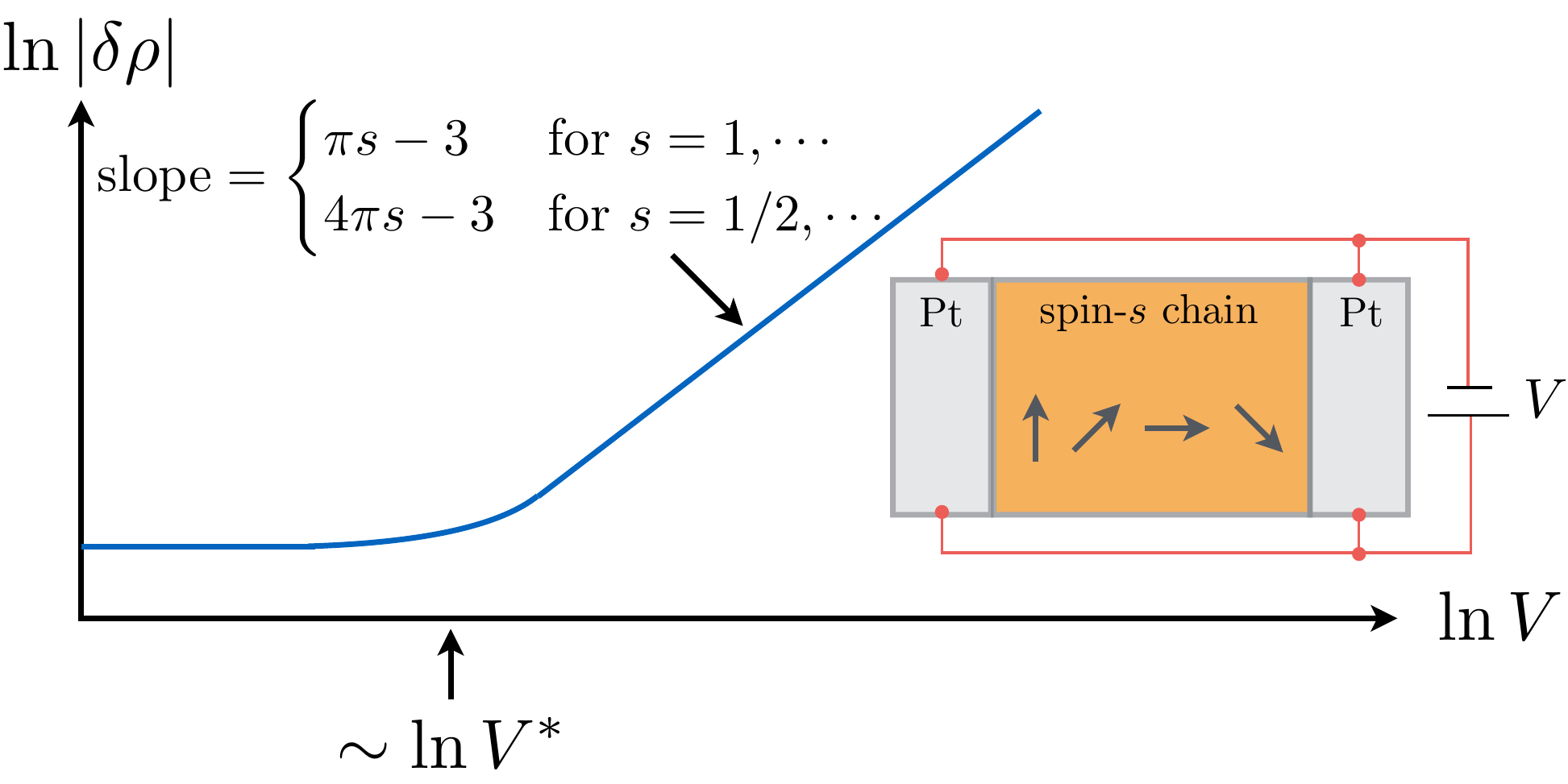}
\caption{(color online) A change in electrical resistance $|\delta \rho|$ of the magnetoelectric circuit as a function of an applied voltage $V$ in logarithmic scale. See the main text for detailed discussions.}
\label{fig:fig3}
\end{figure}

\emph{Experimental proposal.}|The dependence of the decaying rate on constituent spins can be experimentally inferred by measuring electrical resistance of the magnetoelectric circuit that has been proposed for probing superfluid spin transport \cite{TakeiPRL2015}. The circuit consists of a quasi one-dimensional easy-plane antiferromagnet and two parallel-connected metals with strong spin-orbit coupling (e.g., platinum) sandwiching it. See Fig.~\ref{fig:fig3} for schematics of the setup. With charge current flowing, two interfaces of the antiferromagnet to the metals act as a spin source and drain for spin transport via spin-transfer torque. Spin supercurrent is sustained by spiraling texture of the local order parameter within the easy plane. QPS disturb the texture and unwind it by $2 \pi$ for integer $s$ and $4 \pi$ for half-odd-integer $s$ with the frequency $\kappa$. This unwinding of the phase propagates to the ends of spin chains and induce dynamics of spins at the interfaces. Via spin pumping \cite{Brataas2012}, spin rotations generate an electromotive force on electrons in the metals, which decreases the effective resistance of the circuit. 

Following derivations of Refs.~\cite{TakeiPRL2015, KimarXiv2015}, we can calculate the change of the effective resistance: $\rho \to \rho + \delta \rho$ with $\delta \rho = - \vartheta^2 \kappa(I, T) L A / 2 J s^2 d$ (considering QPS as perturbation to uniform spin-current states), where $I$ is the spin current flowing through a single chain, $A$ is the cross section of the antiferromagnet, $\rho$ is the resistivity of the metal, and $\vartheta$ is related to the effective interfacial spin Hall angle $\Theta$ via $\vartheta \equiv (\hbar / 2 e t) \tan \Theta$, with $- e$ being the electric charge of a single electron and $t$ being the thickness of the metals in the direction perpendicular to the interface. Figure~\ref{fig:fig3} schematically depicts the resistance change $\delta \rho$ as a function of a voltage $V$ in logarithmic scale at a fixed temperature. Above the transition voltage $V^*$, at which the spin current is equal to the temperature $I = T$, $\ln |\delta \rho|$ increases linearly as $\ln V$ increases with the slope $2 \mu - 3$ that is determined by constituent spins. Below the transition voltage, $\delta \rho$ converges to a constant value that is determined by the ambient temperature.  

For quantitative estimates, let us take the following parameters of quasi one-dimensional antiferromagnet (CH$_3$)$_4$NMnCl$_3$ \cite{HutchingsPRB1972}: $s = 5/2$, $J s^2 = 85$ K, $J s^2 (a + b) = 2$ K, and $d = 3$ nm. The associated continuum parameters are $\lambda = 10$ nm and $c = 3 \times 10^5$ m/s, which yield the critical spin current $I_c = J s^2 d / \lambda = 18$ K and the crossover temperature $T^* = 5$ K. For geometry of the materials, we consider the platinum metals with thickness $t = 5$ nm and the antiferromagnet with length $L = 1$ $\mu$m and cross section $A = 400$ nm$^2$. Using $\Theta = 0.03$ for the interfacial spin Hall angle (measured for Pt\textbar YIG interfaces \cite{HahnPRB2013}), the change in the effective resistance is $\delta \rho = - 0.5$ $\mu \Omega$ at the spin current of $I = I_c / 10$ and the temperature $T = 3$ K.

\emph{Discussion}|In certain spin chains, dimerization of sites can occur at low temperatures, e.g., as a result of the spin-Peierls transition \cite{BrayPRL1975}. The Hamiltonian then acquires a new term that breaks the sublattice symmetry; $H \to H + \alpha J \sum_{i} (-1)^i \mathbf{S}_i \cdot \mathbf{S}_{i+1}$. The topological term in the nonlinear sigma model changes as well: $\theta = 2 \pi S (1 + \alpha)$ \cite{AffleckPRL1986-2}. With this change of $\theta$, for half-odd-integer $s$, a pair of QPS with skyrmion charges $Q = \pm 1/2$ contributes to the partition function with the prefactor $4 \sin^2 (\pi \alpha / 2)$, which may change the qualitative behavior of the decaying rate of the spin current.

We would like to mention that QPS in topological superconductors occur in multiples of $4 \pi$ (instead of $2 \pi$ in conventional superconductors) \cite{PekkerPRB2013} as in superfluid spin transport through half-odd-integer spin chains.

\begin{acknowledgments}
We are grateful to Ian Affleck for leading our attention to the topological term in the Heisenberg antiferromagnetic spin chains and to Gil Rafael, So Takei, Oleg Tchernyshyov, and Ricardo Zarzuela for insightful discussions. This work was supported by the Army Research Office under Contract No. 911NF-14-1-0016.
\end{acknowledgments}

\bibliographystyle{/Users/evol/Dropbox/School/Research/apsrev4-1-nourl}
\bibliography{/Users/evol/Dropbox/School/Research/master}

\begin{thebibliography}{51}%
\makeatletter
\providecommand \@ifxundefined [1]{%
 \@ifx{#1\undefined}
}%
\providecommand \@ifnum [1]{%
 \ifnum #1\expandafter \@firstoftwo
 \else \expandafter \@secondoftwo
 \fi
}%
\providecommand \@ifx [1]{%
 \ifx #1\expandafter \@firstoftwo
 \else \expandafter \@secondoftwo
 \fi
}%
\providecommand \natexlab [1]{#1}%
\providecommand \enquote  [1]{``#1''}%
\providecommand \bibnamefont  [1]{#1}%
\providecommand \bibfnamefont [1]{#1}%
\providecommand \citenamefont [1]{#1}%
\providecommand \href@noop [0]{\@secondoftwo}%
\providecommand \href [0]{\begingroup \@sanitize@url \@href}%
\providecommand \@href[1]{\@@startlink{#1}\@@href}%
\providecommand \@@href[1]{\endgroup#1\@@endlink}%
\providecommand \@sanitize@url [0]{\catcode `\\12\catcode `\$12\catcode
  `\&12\catcode `\#12\catcode `\^12\catcode `\_12\catcode `\%12\relax}%
\providecommand \@@startlink[1]{}%
\providecommand \@@endlink[0]{}%
\providecommand \url  [0]{\begingroup\@sanitize@url \@url }%
\providecommand \@url [1]{\endgroup\@href {#1}{\urlprefix }}%
\providecommand \urlprefix  [0]{URL }%
\providecommand \Eprint [0]{\href }%
\providecommand \doibase [0]{http://dx.doi.org/}%
\providecommand \selectlanguage [0]{\@gobble}%
\providecommand \bibinfo  [0]{\@secondoftwo}%
\providecommand \bibfield  [0]{\@secondoftwo}%
\providecommand \translation [1]{[#1]}%
\providecommand \BibitemOpen [0]{}%
\providecommand \bibitemStop [0]{}%
\providecommand \bibitemNoStop [0]{.\EOS\space}%
\providecommand \EOS [0]{\spacefactor3000\relax}%
\providecommand \BibitemShut  [1]{\csname bibitem#1\endcsname}%
\let\auto@bib@innerbib\@empty
\bibitem [{\citenamefont {Mermin}\ and\ \citenamefont
  {Wagner}(1966)}]{MerminPRL1966}%
  \BibitemOpen
  \bibfield  {author} {\bibinfo {author} {\bibfnamefont {N.~D.}\ \bibnamefont
  {Mermin}}\ and\ \bibinfo {author} {\bibfnamefont {H.}~\bibnamefont
  {Wagner}},\ }\href {\doibase 10.1103/PhysRevLett.17.1133} {\bibfield
  {journal} {\bibinfo  {journal} {Phys. Rev. Lett.}\ }\textbf {\bibinfo
  {volume} {17}},\ \bibinfo {pages} {1133} (\bibinfo {year}
  {1966})}\BibitemShut {NoStop}%
\bibitem [{\citenamefont {Affleck}(1989)}]{AffleckJPCM1989}%
  \BibitemOpen
  \bibfield  {author} {\bibinfo {author} {\bibfnamefont {I.}~\bibnamefont
  {Affleck}},\ }\href {http://stacks.iop.org/0953-8984/1/i=19/a=001} {\bibfield
   {journal} {\bibinfo  {journal} {J. Phys.: Condens. Matter}\ }\textbf
  {\bibinfo {volume} {1}},\ \bibinfo {pages} {3047} (\bibinfo {year}
  {1989})}\BibitemShut {NoStop}%
\bibitem [{\citenamefont {Mikeska}\ and\ \citenamefont
  {Kolezhuk}(2004)}]{Mikeska2004}%
  \BibitemOpen
  \bibfield  {author} {\bibinfo {author} {\bibfnamefont {H.-J.}\ \bibnamefont
  {Mikeska}}\ and\ \bibinfo {author} {\bibfnamefont {A.}~\bibnamefont
  {Kolezhuk}},\ }\enquote {\bibinfo {title} {One-dimensional magnetism},}\ in\
  \href {http://dx.doi.org/10.1007/BFb0119591} {\emph {\bibinfo {booktitle}
  {Quantum Magnetism}}},\ \bibinfo {editor} {edited by\ \bibinfo {editor}
  {\bibfnamefont {U.}~\bibnamefont {Schollw{\"o}ck}}, \bibinfo {editor}
  {\bibfnamefont {J.}~\bibnamefont {Richter}}, \bibinfo {editor} {\bibfnamefont
  {D.}~\bibnamefont {Farnell}}, \ and\ \bibinfo {editor} {\bibfnamefont
  {R.}~\bibnamefont {Bishop}}}\ (\bibinfo  {publisher} {Springer Berlin
  Heidelberg},\ \bibinfo {year} {2004})\BibitemShut {NoStop}%
\bibitem [{\citenamefont {Haldane}(1983{\natexlab{a}})}]{HaldanePL1983}%
  \BibitemOpen
  \bibfield  {author} {\bibinfo {author} {\bibfnamefont {F.~D.~M.}\
  \bibnamefont {Haldane}},\ }\href {\doibase
  http://dx.doi.org/10.1016/0375-9601(83)90631-X} {\bibfield  {journal}
  {\bibinfo  {journal} {Phys. Lett.}\ }\textbf {\bibinfo {volume} {93A}},\
  \bibinfo {pages} {464 } (\bibinfo {year} {1983}{\natexlab{a}})}\BibitemShut
  {NoStop}%
\bibitem [{\citenamefont {Haldane}(1983{\natexlab{b}})}]{HaldanePRL1983}%
  \BibitemOpen
  \bibfield  {author} {\bibinfo {author} {\bibfnamefont {F.~D.~M.}\
  \bibnamefont {Haldane}},\ }\href {\doibase 10.1103/PhysRevLett.50.1153}
  {\bibfield  {journal} {\bibinfo  {journal} {Phys. Rev. Lett.}\ }\textbf
  {\bibinfo {volume} {50}},\ \bibinfo {pages} {1153} (\bibinfo {year}
  {1983}{\natexlab{b}})}\BibitemShut {NoStop}%
\bibitem [{\citenamefont {Lieb}\ \emph {et~al.}(1961)\citenamefont {Lieb},
  \citenamefont {Schultz},\ and\ \citenamefont {Mattis}}]{LiebAP1961}%
  \BibitemOpen
  \bibfield  {author} {\bibinfo {author} {\bibfnamefont {E.}~\bibnamefont
  {Lieb}}, \bibinfo {author} {\bibfnamefont {T.}~\bibnamefont {Schultz}}, \
  and\ \bibinfo {author} {\bibfnamefont {D.}~\bibnamefont {Mattis}},\ }\href
  {\doibase http://dx.doi.org/10.1016/0003-4916(61)90115-4} {\bibfield
  {journal} {\bibinfo  {journal} {Ann. Phys.}\ }\textbf {\bibinfo {volume}
  {16}},\ \bibinfo {pages} {407 } (\bibinfo {year} {1961})}\BibitemShut
  {NoStop}%
\bibitem [{\citenamefont {Affleck}\ and\ \citenamefont
  {Lieb}(1986)}]{AffleckLMP1986}%
  \BibitemOpen
  \bibfield  {author} {\bibinfo {author} {\bibfnamefont {I.}~\bibnamefont
  {Affleck}}\ and\ \bibinfo {author} {\bibfnamefont {E.}~\bibnamefont {Lieb}},\
  }\href {http://dx.doi.org/10.1007/BF00400304} {\bibfield  {journal} {\bibinfo
   {journal} {Lett. Math. Phys.}\ }\textbf {\bibinfo {volume} {12}},\ \bibinfo
  {pages} {57} (\bibinfo {year} {1986})}\BibitemShut {NoStop}%
\bibitem [{\citenamefont {Shankar}\ and\ \citenamefont
  {Read}(1990)}]{ShankarNPB1990}%
  \BibitemOpen
  \bibfield  {author} {\bibinfo {author} {\bibfnamefont {R.}~\bibnamefont
  {Shankar}}\ and\ \bibinfo {author} {\bibfnamefont {N.}~\bibnamefont {Read}},\
  }\href {\doibase http://dx.doi.org/10.1016/0550-3213(90)90437-I} {\bibfield
  {journal} {\bibinfo  {journal} {Nucl. Phys. B}\ }\textbf {\bibinfo {volume}
  {336}},\ \bibinfo {pages} {457 } (\bibinfo {year} {1990})}\BibitemShut
  {NoStop}%
\bibitem [{\citenamefont {Buyers}\ \emph {et~al.}(1986)\citenamefont {Buyers},
  \citenamefont {Morra}, \citenamefont {Armstrong}, \citenamefont {Hogan},
  \citenamefont {Gerlach},\ and\ \citenamefont {Hirakawa}}]{BuyersPRL1986}%
  \BibitemOpen
  \bibfield  {author} {\bibinfo {author} {\bibfnamefont {W.~J.~L.}\
  \bibnamefont {Buyers}}, \bibinfo {author} {\bibfnamefont {R.~M.}\
  \bibnamefont {Morra}}, \bibinfo {author} {\bibfnamefont {R.~L.}\ \bibnamefont
  {Armstrong}}, \bibinfo {author} {\bibfnamefont {M.~J.}\ \bibnamefont
  {Hogan}}, \bibinfo {author} {\bibfnamefont {P.}~\bibnamefont {Gerlach}}, \
  and\ \bibinfo {author} {\bibfnamefont {K.}~\bibnamefont {Hirakawa}},\ }\href
  {\doibase 10.1103/PhysRevLett.56.371} {\bibfield  {journal} {\bibinfo
  {journal} {Phys. Rev. Lett.}\ }\textbf {\bibinfo {volume} {56}},\ \bibinfo
  {pages} {371} (\bibinfo {year} {1986})}\BibitemShut {NoStop}%
\bibitem [{\citenamefont {Renard}\ \emph {et~al.}(1987)\citenamefont {Renard},
  \citenamefont {Verdaguer}, \citenamefont {Regnault}, \citenamefont
  {Erkelens}, \citenamefont {Rossat-Mignod},\ and\ \citenamefont
  {Stirling}}]{RenardEPL1987}%
  \BibitemOpen
  \bibfield  {author} {\bibinfo {author} {\bibfnamefont {J.~P.}\ \bibnamefont
  {Renard}}, \bibinfo {author} {\bibfnamefont {M.}~\bibnamefont {Verdaguer}},
  \bibinfo {author} {\bibfnamefont {L.~P.}\ \bibnamefont {Regnault}}, \bibinfo
  {author} {\bibfnamefont {W.~A.~C.}\ \bibnamefont {Erkelens}}, \bibinfo
  {author} {\bibfnamefont {J.}~\bibnamefont {Rossat-Mignod}}, \ and\ \bibinfo
  {author} {\bibfnamefont {W.~G.}\ \bibnamefont {Stirling}},\ }\href
  {http://stacks.iop.org/0295-5075/3/i=8/a=013} {\bibfield  {journal} {\bibinfo
   {journal} {Europhys. Lett.}\ }\textbf {\bibinfo {volume} {3}},\ \bibinfo
  {pages} {945} (\bibinfo {year} {1987})}\BibitemShut {NoStop}%
\bibitem [{\citenamefont {Affleck}(1986{\natexlab{a}})}]{AffleckPRL1986-1}%
  \BibitemOpen
  \bibfield  {author} {\bibinfo {author} {\bibfnamefont {I.}~\bibnamefont
  {Affleck}},\ }\href {\doibase 10.1103/PhysRevLett.56.408} {\bibfield
  {journal} {\bibinfo  {journal} {Phys. Rev. Lett.}\ }\textbf {\bibinfo
  {volume} {56}},\ \bibinfo {pages} {408} (\bibinfo {year}
  {1986}{\natexlab{a}})}\BibitemShut {NoStop}%
\bibitem [{\citenamefont {Mikeska}(1980)}]{MikeskaJPC1980}%
  \BibitemOpen
  \bibfield  {author} {\bibinfo {author} {\bibfnamefont {H.~J.}\ \bibnamefont
  {Mikeska}},\ }\href {http://stacks.iop.org/0022-3719/13/i=15/a=015}
  {\bibfield  {journal} {\bibinfo  {journal} {J. Phys. C: Solid State Phys.}\
  }\textbf {\bibinfo {volume} {13}},\ \bibinfo {pages} {2913} (\bibinfo {year}
  {1980})}\BibitemShut {NoStop}%
\bibitem [{\citenamefont {Fradkin}\ and\ \citenamefont
  {Stone}(1988)}]{FradkinPRB1988}%
  \BibitemOpen
  \bibfield  {author} {\bibinfo {author} {\bibfnamefont {E.}~\bibnamefont
  {Fradkin}}\ and\ \bibinfo {author} {\bibfnamefont {M.}~\bibnamefont
  {Stone}},\ }\href {\doibase 10.1103/PhysRevB.38.7215} {\bibfield  {journal}
  {\bibinfo  {journal} {Phys. Rev. B}\ }\textbf {\bibinfo {volume} {38}},\
  \bibinfo {pages} {7215} (\bibinfo {year} {1988})}\BibitemShut {NoStop}%
\bibitem [{\citenamefont {Gross}(1978)}]{GrossNPB1978}%
  \BibitemOpen
  \bibfield  {author} {\bibinfo {author} {\bibfnamefont {D.~J.}\ \bibnamefont
  {Gross}},\ }\href {\doibase http://dx.doi.org/10.1016/0550-3213(78)90470-4}
  {\bibfield  {journal} {\bibinfo  {journal} {Nucl. Phys.}\ }\textbf {\bibinfo
  {volume} {B132}},\ \bibinfo {pages} {439 } (\bibinfo {year}
  {1978})}\BibitemShut {NoStop}%
\bibitem [{\citenamefont {Ivanov}\ \emph {et~al.}(1998)\citenamefont {Ivanov},
  \citenamefont {Kolezhuk},\ and\ \citenamefont {Kireev}}]{IvanovPRB1998}%
  \BibitemOpen
  \bibfield  {author} {\bibinfo {author} {\bibfnamefont {B.~A.}\ \bibnamefont
  {Ivanov}}, \bibinfo {author} {\bibfnamefont {A.~K.}\ \bibnamefont
  {Kolezhuk}}, \ and\ \bibinfo {author} {\bibfnamefont {V.~E.}\ \bibnamefont
  {Kireev}},\ }\href {\doibase 10.1103/PhysRevB.58.11514} {\bibfield  {journal}
  {\bibinfo  {journal} {Phys. Rev. B}\ }\textbf {\bibinfo {volume} {58}},\
  \bibinfo {pages} {11514} (\bibinfo {year} {1998})}\BibitemShut {NoStop}%
\bibitem [{\citenamefont {Sonin}(1978)}]{SoninJETP1978}%
  \BibitemOpen
  \bibfield  {author} {\bibinfo {author} {\bibfnamefont {E.~B.}\ \bibnamefont
  {Sonin}},\ }\href@noop {} {\bibfield  {journal} {\bibinfo  {journal} {JETP
  Lett.}\ }\textbf {\bibinfo {volume} {47}},\ \bibinfo {pages} {1091} (\bibinfo
  {year} {1978})}\BibitemShut {NoStop}%
\bibitem [{\citenamefont {Sonin}(2010)}]{SoninAP2010}%
  \BibitemOpen
  \bibfield  {author} {\bibinfo {author} {\bibfnamefont {E.}~\bibnamefont
  {Sonin}},\ }\href {\doibase 10.1080/00018731003739943} {\bibfield  {journal}
  {\bibinfo  {journal} {Adv. Phys.}\ }\textbf {\bibinfo {volume} {59}},\
  \bibinfo {pages} {181} (\bibinfo {year} {2010})}\BibitemShut {NoStop}%
\bibitem [{\citenamefont {K\"onig}\ \emph {et~al.}(2001)\citenamefont
  {K\"onig}, \citenamefont {B\o{}nsager},\ and\ \citenamefont
  {MacDonald}}]{KonigPRL2002}%
  \BibitemOpen
  \bibfield  {author} {\bibinfo {author} {\bibfnamefont {J.}~\bibnamefont
  {K\"onig}}, \bibinfo {author} {\bibfnamefont {M.~C.}\ \bibnamefont
  {B\o{}nsager}}, \ and\ \bibinfo {author} {\bibfnamefont {A.~H.}\ \bibnamefont
  {MacDonald}},\ }\href {\doibase 10.1103/PhysRevLett.87.187202} {\bibfield
  {journal} {\bibinfo  {journal} {Phys. Rev. Lett.}\ }\textbf {\bibinfo
  {volume} {87}},\ \bibinfo {pages} {187202} (\bibinfo {year}
  {2001})}\BibitemShut {NoStop}%
\bibitem [{\citenamefont {Chen}\ \emph {et~al.}(2014)\citenamefont {Chen},
  \citenamefont {Kent}, \citenamefont {MacDonald},\ and\ \citenamefont
  {Sodemann}}]{ChenPRB2014}%
  \BibitemOpen
  \bibfield  {author} {\bibinfo {author} {\bibfnamefont {H.}~\bibnamefont
  {Chen}}, \bibinfo {author} {\bibfnamefont {A.~D.}\ \bibnamefont {Kent}},
  \bibinfo {author} {\bibfnamefont {A.~H.}\ \bibnamefont {MacDonald}}, \ and\
  \bibinfo {author} {\bibfnamefont {I.}~\bibnamefont {Sodemann}},\ }\href
  {\doibase 10.1103/PhysRevB.90.220401} {\bibfield  {journal} {\bibinfo
  {journal} {Phys. Rev. B}\ }\textbf {\bibinfo {volume} {90}},\ \bibinfo
  {pages} {220401} (\bibinfo {year} {2014})}\BibitemShut {NoStop}%
\bibitem [{\citenamefont {Chen}\ and\ \citenamefont
  {Sigrist}(2014)}]{ChenPRB2014-2}%
  \BibitemOpen
  \bibfield  {author} {\bibinfo {author} {\bibfnamefont {W.}~\bibnamefont
  {Chen}}\ and\ \bibinfo {author} {\bibfnamefont {M.}~\bibnamefont {Sigrist}},\
  }\href {\doibase 10.1103/PhysRevB.89.024511} {\bibfield  {journal} {\bibinfo
  {journal} {Phys. Rev. B}\ }\textbf {\bibinfo {volume} {89}},\ \bibinfo
  {pages} {024511} (\bibinfo {year} {2014})}\BibitemShut {NoStop}%
\bibitem [{\citenamefont {Chen}\ and\ \citenamefont
  {Sigrist}(2015)}]{ChenPRL2015}%
  \BibitemOpen
  \bibfield  {author} {\bibinfo {author} {\bibfnamefont {W.}~\bibnamefont
  {Chen}}\ and\ \bibinfo {author} {\bibfnamefont {M.}~\bibnamefont {Sigrist}},\
  }\href {\doibase 10.1103/PhysRevLett.114.157203} {\bibfield  {journal}
  {\bibinfo  {journal} {Phys. Rev. Lett.}\ }\textbf {\bibinfo {volume} {114}},\
  \bibinfo {pages} {157203} (\bibinfo {year} {2015})}\BibitemShut {NoStop}%
\bibitem [{\citenamefont {Takei}\ and\ \citenamefont
  {Tserkovnyak}(2014)}]{TakeiPRL2014}%
  \BibitemOpen
  \bibfield  {author} {\bibinfo {author} {\bibfnamefont {S.}~\bibnamefont
  {Takei}}\ and\ \bibinfo {author} {\bibfnamefont {Y.}~\bibnamefont
  {Tserkovnyak}},\ }\href {\doibase 10.1103/PhysRevLett.112.227201} {\bibfield
  {journal} {\bibinfo  {journal} {Phys. Rev. Lett.}\ }\textbf {\bibinfo
  {volume} {112}},\ \bibinfo {pages} {227201} (\bibinfo {year}
  {2014})}\BibitemShut {NoStop}%
\bibitem [{\citenamefont {Takei}\ \emph {et~al.}(2014)\citenamefont {Takei},
  \citenamefont {Halperin}, \citenamefont {Yacoby},\ and\ \citenamefont
  {Tserkovnyak}}]{TakeiPRB2014}%
  \BibitemOpen
  \bibfield  {author} {\bibinfo {author} {\bibfnamefont {S.}~\bibnamefont
  {Takei}}, \bibinfo {author} {\bibfnamefont {B.~I.}\ \bibnamefont {Halperin}},
  \bibinfo {author} {\bibfnamefont {A.}~\bibnamefont {Yacoby}}, \ and\ \bibinfo
  {author} {\bibfnamefont {Y.}~\bibnamefont {Tserkovnyak}},\ }\href {\doibase
  10.1103/PhysRevB.90.094408} {\bibfield  {journal} {\bibinfo  {journal} {Phys.
  Rev. B}\ }\textbf {\bibinfo {volume} {90}},\ \bibinfo {pages} {094408}
  (\bibinfo {year} {2014})}\BibitemShut {NoStop}%
\bibitem [{\citenamefont {Halperin}\ \emph {et~al.}(2010)\citenamefont
  {Halperin}, \citenamefont {Refael},\ and\ \citenamefont
  {Demler}}]{HalperinIJMPB2010}%
  \BibitemOpen
  \bibfield  {author} {\bibinfo {author} {\bibfnamefont {B.~I.}\ \bibnamefont
  {Halperin}}, \bibinfo {author} {\bibfnamefont {G.}~\bibnamefont {Refael}}, \
  and\ \bibinfo {author} {\bibfnamefont {E.}~\bibnamefont {Demler}},\ }\href
  {\doibase 10.1142/S021797921005644X} {\bibfield  {journal} {\bibinfo
  {journal} {Int. J. Mod. Phys. B}\ }\textbf {\bibinfo {volume} {24}},\
  \bibinfo {pages} {4039} (\bibinfo {year} {2010})}\BibitemShut {NoStop}%
\bibitem [{\citenamefont {Little}(1967)}]{LittlePR1967}%
  \BibitemOpen
  \bibfield  {author} {\bibinfo {author} {\bibfnamefont {W.~A.}\ \bibnamefont
  {Little}},\ }\href {\doibase 10.1103/PhysRev.156.396} {\bibfield  {journal}
  {\bibinfo  {journal} {Phys. Rev.}\ }\textbf {\bibinfo {volume} {156}},\
  \bibinfo {pages} {396} (\bibinfo {year} {1967})}\BibitemShut {NoStop}%
\bibitem [{\citenamefont {Langer}\ and\ \citenamefont
  {Ambegaokar}(1967)}]{LangerPR1967}%
  \BibitemOpen
  \bibfield  {author} {\bibinfo {author} {\bibfnamefont {J.~S.}\ \bibnamefont
  {Langer}}\ and\ \bibinfo {author} {\bibfnamefont {V.}~\bibnamefont
  {Ambegaokar}},\ }\href {\doibase 10.1103/PhysRev.164.498} {\bibfield
  {journal} {\bibinfo  {journal} {Phys. Rev.}\ }\textbf {\bibinfo {volume}
  {164}},\ \bibinfo {pages} {498} (\bibinfo {year} {1967})}\BibitemShut
  {NoStop}%
\bibitem [{\citenamefont {McCumber}\ and\ \citenamefont
  {Halperin}(1970)}]{McCumberPRB1970}%
  \BibitemOpen
  \bibfield  {author} {\bibinfo {author} {\bibfnamefont {D.~E.}\ \bibnamefont
  {McCumber}}\ and\ \bibinfo {author} {\bibfnamefont {B.~I.}\ \bibnamefont
  {Halperin}},\ }\href {\doibase 10.1103/PhysRevB.1.1054} {\bibfield  {journal}
  {\bibinfo  {journal} {Phys. Rev. B}\ }\textbf {\bibinfo {volume} {1}},\
  \bibinfo {pages} {1054} (\bibinfo {year} {1970})}\BibitemShut {NoStop}%
\bibitem [{\citenamefont {Kim}\ \emph {et~al.}()\citenamefont {Kim},
  \citenamefont {Takei},\ and\ \citenamefont {Tserkovnyak}}]{KimarXiv2015}%
  \BibitemOpen
  \bibfield  {author} {\bibinfo {author} {\bibfnamefont {S.~K.}\ \bibnamefont
  {Kim}}, \bibinfo {author} {\bibfnamefont {S.}~\bibnamefont {Takei}}, \ and\
  \bibinfo {author} {\bibfnamefont {Y.}~\bibnamefont {Tserkovnyak}},\
  }\href@noop {} {}\Eprint {http://arxiv.org/abs/1509.00904} {arXiv:1509.00904}
  \BibitemShut {NoStop}%
\bibitem [{\citenamefont {Giodano}(1994)}]{GiordanoP1994}%
  \BibitemOpen
  \bibfield  {author} {\bibinfo {author} {\bibfnamefont {N.}~\bibnamefont
  {Giodano}},\ }\href {\doibase http://dx.doi.org/10.1016/0921-4526(94)90097-3}
  {\bibfield  {journal} {\bibinfo  {journal} {Physica (Amsterdam)}\ }\textbf
  {\bibinfo {volume} {203B}},\ \bibinfo {pages} {460 } (\bibinfo {year}
  {1994})}\BibitemShut {NoStop}%
\bibitem [{\citenamefont {Zaikin}\ \emph {et~al.}(1997)\citenamefont {Zaikin},
  \citenamefont {Golubev}, \citenamefont {van Otterlo},\ and\ \citenamefont
  {Zim\'anyi}}]{ZaikinPRL1997}%
  \BibitemOpen
  \bibfield  {author} {\bibinfo {author} {\bibfnamefont {A.~D.}\ \bibnamefont
  {Zaikin}}, \bibinfo {author} {\bibfnamefont {D.~S.}\ \bibnamefont {Golubev}},
  \bibinfo {author} {\bibfnamefont {A.}~\bibnamefont {van Otterlo}}, \ and\
  \bibinfo {author} {\bibfnamefont {G.~T.}\ \bibnamefont {Zim\'anyi}},\ }\href
  {\doibase 10.1103/PhysRevLett.78.1552} {\bibfield  {journal} {\bibinfo
  {journal} {Phys. Rev. Lett.}\ }\textbf {\bibinfo {volume} {78}},\ \bibinfo
  {pages} {1552} (\bibinfo {year} {1997})}\BibitemShut {NoStop}%
\bibitem [{\citenamefont {Takei}\ and\ \citenamefont
  {Tserkovnyak}(2015)}]{TakeiPRL2015}%
  \BibitemOpen
  \bibfield  {author} {\bibinfo {author} {\bibfnamefont {S.}~\bibnamefont
  {Takei}}\ and\ \bibinfo {author} {\bibfnamefont {Y.}~\bibnamefont
  {Tserkovnyak}},\ }\href {\doibase 10.1103/PhysRevLett.115.156604} {\bibfield
  {journal} {\bibinfo  {journal} {Phys. Rev. Lett.}\ }\textbf {\bibinfo
  {volume} {115}},\ \bibinfo {pages} {156604} (\bibinfo {year}
  {2015})}\BibitemShut {NoStop}%
\bibitem [{\citenamefont {Hutchings}\ \emph {et~al.}(1972)\citenamefont
  {Hutchings}, \citenamefont {Shirane}, \citenamefont {Birgeneau},\ and\
  \citenamefont {Holt}}]{HutchingsPRB1972}%
  \BibitemOpen
  \bibfield  {author} {\bibinfo {author} {\bibfnamefont {M.~T.}\ \bibnamefont
  {Hutchings}}, \bibinfo {author} {\bibfnamefont {G.}~\bibnamefont {Shirane}},
  \bibinfo {author} {\bibfnamefont {R.~J.}\ \bibnamefont {Birgeneau}}, \ and\
  \bibinfo {author} {\bibfnamefont {S.~L.}\ \bibnamefont {Holt}},\ }\href
  {\doibase 10.1103/PhysRevB.5.1999} {\bibfield  {journal} {\bibinfo  {journal}
  {Phys. Rev. B}\ }\textbf {\bibinfo {volume} {5}},\ \bibinfo {pages} {1999}
  (\bibinfo {year} {1972})}\BibitemShut {NoStop}%
\bibitem [{\citenamefont {Boucher}\ \emph {et~al.}(1979)\citenamefont
  {Boucher}, \citenamefont {Regnault}, \citenamefont {Rossat-Mignod},
  \citenamefont {Villain},\ and\ \citenamefont {Renard}}]{BoucherJMMM1979}%
  \BibitemOpen
  \bibfield  {author} {\bibinfo {author} {\bibfnamefont {J.}~\bibnamefont
  {Boucher}}, \bibinfo {author} {\bibfnamefont {L.}~\bibnamefont {Regnault}},
  \bibinfo {author} {\bibfnamefont {J.}~\bibnamefont {Rossat-Mignod}}, \bibinfo
  {author} {\bibfnamefont {J.}~\bibnamefont {Villain}}, \ and\ \bibinfo
  {author} {\bibfnamefont {J.}~\bibnamefont {Renard}},\ }\href {\doibase
  http://dx.doi.org/10.1016/0304-8853(79)90105-7} {\bibfield  {journal}
  {\bibinfo  {journal} {J. Magn. Magn. Mater.}\ }\textbf {\bibinfo {volume}
  {14}},\ \bibinfo {pages} {155 } (\bibinfo {year} {1979})}\BibitemShut
  {NoStop}%
\bibitem [{\citenamefont {Fl{\"u}ggen}\ and\ \citenamefont
  {Mikeska}(1983)}]{FluggenSSC1983}%
  \BibitemOpen
  \bibfield  {author} {\bibinfo {author} {\bibfnamefont {N.}~\bibnamefont
  {Fl{\"u}ggen}}\ and\ \bibinfo {author} {\bibfnamefont {H.}~\bibnamefont
  {Mikeska}},\ }\href {\doibase http://dx.doi.org/10.1016/0038-1098(83)90290-9}
  {\bibfield  {journal} {\bibinfo  {journal} {Solid State Commun.}\ }\textbf
  {\bibinfo {volume} {48}},\ \bibinfo {pages} {293 } (\bibinfo {year}
  {1983})}\BibitemShut {NoStop}%
\bibitem [{\citenamefont {Ivanov}\ and\ \citenamefont
  {Sheka}(1994)}]{IvanovPRL1994}%
  \BibitemOpen
  \bibfield  {author} {\bibinfo {author} {\bibfnamefont {B.~A.}\ \bibnamefont
  {Ivanov}}\ and\ \bibinfo {author} {\bibfnamefont {D.~D.}\ \bibnamefont
  {Sheka}},\ }\href {\doibase 10.1103/PhysRevLett.72.404} {\bibfield  {journal}
  {\bibinfo  {journal} {Phys. Rev. Lett.}\ }\textbf {\bibinfo {volume} {72}},\
  \bibinfo {pages} {404} (\bibinfo {year} {1994})}\BibitemShut {NoStop}%
\bibitem [{\citenamefont {Tretiakov}\ and\ \citenamefont
  {Tchernyshyov}(2007)}]{TretiakovPRB2007}%
  \BibitemOpen
  \bibfield  {author} {\bibinfo {author} {\bibfnamefont {O.~A.}\ \bibnamefont
  {Tretiakov}}\ and\ \bibinfo {author} {\bibfnamefont {O.}~\bibnamefont
  {Tchernyshyov}},\ }\href {\doibase 10.1103/PhysRevB.75.012408} {\bibfield
  {journal} {\bibinfo  {journal} {Phys. Rev. B}\ }\textbf {\bibinfo {volume}
  {75}},\ \bibinfo {pages} {012408} (\bibinfo {year} {2007})}\BibitemShut
  {NoStop}%
\bibitem [{Note1()}]{Note1}%
  \BibitemOpen
  \bibinfo {note} {With the background spin current $k$, $S_\protect \text
  {core}$ acquires a new contribution that is order of $k^2$. This change of
  $S_\protect \text {core}$ does not affect qualitative behavior of QPS, and is
  thus ignored.}\BibitemShut {Stop}%
\bibitem [{\citenamefont {Kosterlitz}\ and\ \citenamefont
  {Thouless}(1973)}]{KosterlitzJPC1973}%
  \BibitemOpen
  \bibfield  {author} {\bibinfo {author} {\bibfnamefont {J.~M.}\ \bibnamefont
  {Kosterlitz}}\ and\ \bibinfo {author} {\bibfnamefont {D.~J.}\ \bibnamefont
  {Thouless}},\ }\href {http://stacks.iop.org/0022-3719/6/i=7/a=010} {\bibfield
   {journal} {\bibinfo  {journal} {J. Phys. C: Solid State Phys.}\ }\textbf
  {\bibinfo {volume} {6}},\ \bibinfo {pages} {1181} (\bibinfo {year}
  {1973})}\BibitemShut {NoStop}%
\bibitem [{\citenamefont {Kosterlitz}(1974)}]{KosterlitzJPC1974}%
  \BibitemOpen
  \bibfield  {author} {\bibinfo {author} {\bibfnamefont {J.~M.}\ \bibnamefont
  {Kosterlitz}},\ }\href {http://stacks.iop.org/0022-3719/7/i=6/a=005}
  {\bibfield  {journal} {\bibinfo  {journal} {J. Phys. C: Solid State Phys.}\
  }\textbf {\bibinfo {volume} {7}},\ \bibinfo {pages} {1046} (\bibinfo {year}
  {1974})}\BibitemShut {NoStop}%
\bibitem [{Note2()}]{Note2}%
  \BibitemOpen
  \bibinfo {note} {To keep the discussion on the phase transition simple, we
  used the bare value of the critical strength, $\mu ^* \approx 2$, which does
  not take its renormalization into account.}\BibitemShut {Stop}%
\bibitem [{\citenamefont {Langer}(1967)}]{LangerAP1967}%
  \BibitemOpen
  \bibfield  {author} {\bibinfo {author} {\bibfnamefont {J.}~\bibnamefont
  {Langer}},\ }\href {\doibase http://dx.doi.org/10.1016/0003-4916(67)90200-X}
  {\bibfield  {journal} {\bibinfo  {journal} {Ann. Phys.}\ }\textbf {\bibinfo
  {volume} {41}},\ \bibinfo {pages} {108 } (\bibinfo {year}
  {1967})}\BibitemShut {NoStop}%
\bibitem [{\citenamefont {Weiss}\ \emph {et~al.}(1987)\citenamefont {Weiss},
  \citenamefont {Grabert}, \citenamefont {H\"anggi},\ and\ \citenamefont
  {Riseborough}}]{WeissPRB1987}%
  \BibitemOpen
  \bibfield  {author} {\bibinfo {author} {\bibfnamefont {U.}~\bibnamefont
  {Weiss}}, \bibinfo {author} {\bibfnamefont {H.}~\bibnamefont {Grabert}},
  \bibinfo {author} {\bibfnamefont {P.}~\bibnamefont {H\"anggi}}, \ and\
  \bibinfo {author} {\bibfnamefont {P.}~\bibnamefont {Riseborough}},\ }\href
  {\doibase 10.1103/PhysRevB.35.9535} {\bibfield  {journal} {\bibinfo
  {journal} {Phys. Rev. B}\ }\textbf {\bibinfo {volume} {35}},\ \bibinfo
  {pages} {9535} (\bibinfo {year} {1987})}\BibitemShut {NoStop}%
\bibitem [{Note3()}]{Note3}%
  \BibitemOpen
  \bibinfo {note} {Equation~(\ref {eq:kappa1}) for $\kappa $ is obtained
  without taking into account quantum fluctuations about vortex solutions,
  which may add an additional dimensionless factor that depends neither on $I$
  nor $T$ \cite {ZaikinPRL1997}.}\BibitemShut {Stop}%
\bibitem [{Note4()}]{Note4}%
  \BibitemOpen
  \bibinfo {note} {There are a few distinct ways to obtain the resistivity of
  one-dimensional superfluid (see e.g., Refs.~\cite {GiamarchiPRB1992,
  DanshitaPRA2012}), which yield different expressions for the decaying rate
  $\kappa (I, T)$. In two limits, $I \ll T$ and $T \ll I$, however, the
  dependence of $\kappa (I, T)$ on the spin current $I$ and the temperature $T$
  is independent of employed methods.}\BibitemShut {Stop}%
\bibitem [{\citenamefont {Brataas}\ \emph {et~al.}(2012)\citenamefont
  {Brataas}, \citenamefont {Tserkovnyak}, \citenamefont {Bauer},\ and\
  \citenamefont {Kelly}}]{Brataas2012}%
  \BibitemOpen
  \bibfield  {author} {\bibinfo {author} {\bibfnamefont {A.}~\bibnamefont
  {Brataas}}, \bibinfo {author} {\bibfnamefont {Y.}~\bibnamefont
  {Tserkovnyak}}, \bibinfo {author} {\bibfnamefont {G.~E.~W.}\ \bibnamefont
  {Bauer}}, \ and\ \bibinfo {author} {\bibfnamefont {P.~J.}\ \bibnamefont
  {Kelly}},\ }in\ \href@noop {} {\emph {\bibinfo {booktitle} {Spin Current}}},\
  \bibinfo {editor} {edited by\ \bibinfo {editor} {\bibfnamefont
  {S.}~\bibnamefont {Maekawa}}, \bibinfo {editor} {\bibfnamefont {S.~O.}\
  \bibnamefont {Valenzuela}}, \bibinfo {editor} {\bibfnamefont
  {E.}~\bibnamefont {Saitoh}}, \ and\ \bibinfo {editor} {\bibfnamefont
  {T.}~\bibnamefont {Kimura}}}\ (\bibinfo  {publisher} {Oxford University
  Press},\ \bibinfo {year} {2012})\BibitemShut {NoStop}%
\bibitem [{\citenamefont {Hahn}\ \emph {et~al.}(2013)\citenamefont {Hahn},
  \citenamefont {de~Loubens}, \citenamefont {Klein}, \citenamefont {Viret},
  \citenamefont {Naletov},\ and\ \citenamefont {Ben~Youssef}}]{HahnPRB2013}%
  \BibitemOpen
  \bibfield  {author} {\bibinfo {author} {\bibfnamefont {C.}~\bibnamefont
  {Hahn}}, \bibinfo {author} {\bibfnamefont {G.}~\bibnamefont {de~Loubens}},
  \bibinfo {author} {\bibfnamefont {O.}~\bibnamefont {Klein}}, \bibinfo
  {author} {\bibfnamefont {M.}~\bibnamefont {Viret}}, \bibinfo {author}
  {\bibfnamefont {V.~V.}\ \bibnamefont {Naletov}}, \ and\ \bibinfo {author}
  {\bibfnamefont {J.}~\bibnamefont {Ben~Youssef}},\ }\href {\doibase
  10.1103/PhysRevB.87.174417} {\bibfield  {journal} {\bibinfo  {journal} {Phys.
  Rev. B}\ }\textbf {\bibinfo {volume} {87}},\ \bibinfo {pages} {174417}
  (\bibinfo {year} {2013})}\BibitemShut {NoStop}%
\bibitem [{\citenamefont {Bray}\ \emph {et~al.}(1975)\citenamefont {Bray},
  \citenamefont {Hart}, \citenamefont {Interrante}, \citenamefont {Jacobs},
  \citenamefont {Kasper}, \citenamefont {Watkins}, \citenamefont {Wee},\ and\
  \citenamefont {Bonner}}]{BrayPRL1975}%
  \BibitemOpen
  \bibfield  {author} {\bibinfo {author} {\bibfnamefont {J.~W.}\ \bibnamefont
  {Bray}}, \bibinfo {author} {\bibfnamefont {H.~R.}\ \bibnamefont {Hart}},
  \bibinfo {author} {\bibfnamefont {L.~V.}\ \bibnamefont {Interrante}},
  \bibinfo {author} {\bibfnamefont {I.~S.}\ \bibnamefont {Jacobs}}, \bibinfo
  {author} {\bibfnamefont {J.~S.}\ \bibnamefont {Kasper}}, \bibinfo {author}
  {\bibfnamefont {G.~D.}\ \bibnamefont {Watkins}}, \bibinfo {author}
  {\bibfnamefont {S.~H.}\ \bibnamefont {Wee}}, \ and\ \bibinfo {author}
  {\bibfnamefont {J.~C.}\ \bibnamefont {Bonner}},\ }\href {\doibase
  10.1103/PhysRevLett.35.744} {\bibfield  {journal} {\bibinfo  {journal} {Phys.
  Rev. Lett.}\ }\textbf {\bibinfo {volume} {35}},\ \bibinfo {pages} {744}
  (\bibinfo {year} {1975})}\BibitemShut {NoStop}%
\bibitem [{\citenamefont {Affleck}(1986{\natexlab{b}})}]{AffleckPRL1986-2}%
  \BibitemOpen
  \bibfield  {author} {\bibinfo {author} {\bibfnamefont {I.}~\bibnamefont
  {Affleck}},\ }\href {\doibase 10.1103/PhysRevLett.57.1048} {\bibfield
  {journal} {\bibinfo  {journal} {Phys. Rev. Lett.}\ }\textbf {\bibinfo
  {volume} {57}},\ \bibinfo {pages} {1048} (\bibinfo {year}
  {1986}{\natexlab{b}})}\BibitemShut {NoStop}%
\bibitem [{\citenamefont {Pekker}\ \emph {et~al.}(2013)\citenamefont {Pekker},
  \citenamefont {Hou}, \citenamefont {Bergman}, \citenamefont {Goldberg},
  \citenamefont {Adagideli},\ and\ \citenamefont {Hassler}}]{PekkerPRB2013}%
  \BibitemOpen
  \bibfield  {author} {\bibinfo {author} {\bibfnamefont {D.}~\bibnamefont
  {Pekker}}, \bibinfo {author} {\bibfnamefont {C.-Y.}\ \bibnamefont {Hou}},
  \bibinfo {author} {\bibfnamefont {D.~L.}\ \bibnamefont {Bergman}}, \bibinfo
  {author} {\bibfnamefont {S.}~\bibnamefont {Goldberg}}, \bibinfo {author}
  {\bibfnamefont {{\.I}.}~\bibnamefont {Adagideli}}, \ and\ \bibinfo {author}
  {\bibfnamefont {F.}~\bibnamefont {Hassler}},\ }\href {\doibase
  10.1103/PhysRevB.87.064506} {\bibfield  {journal} {\bibinfo  {journal} {Phys.
  Rev. B}\ }\textbf {\bibinfo {volume} {87}},\ \bibinfo {pages} {064506}
  (\bibinfo {year} {2013})}\BibitemShut {NoStop}%
\bibitem [{\citenamefont {Giamarchi}(1992)}]{GiamarchiPRB1992}%
  \BibitemOpen
  \bibfield  {author} {\bibinfo {author} {\bibfnamefont {T.}~\bibnamefont
  {Giamarchi}},\ }\href {\doibase 10.1103/PhysRevB.46.342} {\bibfield
  {journal} {\bibinfo  {journal} {Phys. Rev. B}\ }\textbf {\bibinfo {volume}
  {46}},\ \bibinfo {pages} {342} (\bibinfo {year} {1992})}\BibitemShut
  {NoStop}%
\bibitem [{\citenamefont {Danshita}\ and\ \citenamefont
  {Polkovnikov}(2012)}]{DanshitaPRA2012}%
  \BibitemOpen
  \bibfield  {author} {\bibinfo {author} {\bibfnamefont {I.}~\bibnamefont
  {Danshita}}\ and\ \bibinfo {author} {\bibfnamefont {A.}~\bibnamefont
  {Polkovnikov}},\ }\href {\doibase 10.1103/PhysRevA.85.023638} {\bibfield
  {journal} {\bibinfo  {journal} {Phys. Rev. A}\ }\textbf {\bibinfo {volume}
  {85}},\ \bibinfo {pages} {023638} (\bibinfo {year} {2012})}\BibitemShut
  {NoStop}%
\end{thebibliography}%

\end{document}